\documentclass[sigconf]{acmart}
\AtBeginDocument{%
  }

\newcommand{\eg}{\textit{e.g.}}
\newcommand{\ie}{\textit{i.e.}}

\newcommand{\system}{\textsc{ThematicPlane}}

\NewDocumentCommand{\heng}
{ mO{} }{\textcolor{red}{\textsuperscript{\textit{Heng}}\textsf{\textbf{\small[#1]}}}}

\copyrightyear{2025}
\acmYear{2025}
\setcopyright{rightsretained}
\acmConference[UIST Adjunct '25]{The 38th Annual ACM Symposium on User
Interface Software and Technology}{September 28-October 1, 2025}{Busan,
Republic of Korea}
\acmBooktitle{The 38th Annual ACM Symposium on User Interface Software and
Technology (UIST Adjunct '25), September 28-October 1, 2025, Busan, Republic of Korea}
\acmDOI{10.1145/3746058.3758376}
\acmISBN{979-8-4007-2036-9/2025/09}

\begin{document}

\title[\system{}]{\system{}: Bridging Tacit User Intent and Latent Spaces \\ for Image Editing}

\author{Daniel Lee}
\orcid{0009-0005-3061-1827}
\affiliation{%
  \institution{Adobe Inc.}
  \city{San Jose}
  \state{CA}
  \country{USA}}
  \email{dlee1@adobe.com}

\author{Nikhil Sharma}
\orcid{0009-0004-9183-6811}
\affiliation{%
  \institution{Johns Hopkins University}
  \city{Baltimore}
  \state{MD}
  \country{USA}}
  \email{nsharm27@jhu.edu}

\author{Donghoon Shin}
\orcid{0000-0001-9689-7841}
\affiliation{%
  \institution{University of Washington}
  \city{Seattle}
  \state{WA}
  \country{USA}}
  \email{dhoon@uw.edu}

\author{DaEun Choi}
\orcid{0000-0002-4843-0486}
\affiliation{%
  \institution{KAIST}
  \city{Daejeon}
  \country{Republic of Korea}}
  \email{daeun.choi@kaist.ac.kr}

\author{Harsh Sharma}
\orcid{0009-0008-1032-8866}
\affiliation{%
  \institution{University of Colorado}
  \city{Boulder}
  \state{CO}
  \country{USA}}
  \email{harsh.sharma@colorado.edu}
  
\author{Jeonghwan Kim}
\orcid{0009-0002-7277-2168}
\affiliation{%
  \institution{University of Illinois at Urbana-Champaign}
  \city{Champaign}
  \state{IL}
  \country{USA}}
  \email{jk100@illinois.edu}

\author{Heng Ji}
\orcid{0000-0002-0464-7966}
\affiliation{%
  \institution{University of Illinois at Urbana-Champaign}
  \city{Champaign}
  \state{IL}
  \country{USA}}
  \email{hengji@illinois.edu}

\renewcommand{\shortauthors}{Lee, et al.}

\begin{abstract}
Generative AI has made image creation more accessible, yet aligning outputs with nuanced creative intent remains challenging, particularly for non-experts. Existing tools often require users to externalize ideas through prompts or references, limiting fluid exploration. We introduce \system{}, a system that enables users to navigate and manipulate high-level semantic concepts (\eg{}, mood, style, or narrative tone) within an interactive thematic design plane. This interface bridges the gap between tacit creative intent and system control. In our exploratory study ($N=6$), participants engaged in divergent and convergent creative modes, often embracing unexpected results as inspiration or iteration cues. While they grounded their exploration in familiar themes, differing expectations of how themes mapped to outputs revealed a need for more explainable controls. Overall, \system{} fosters expressive, iterative workflows and highlights new directions for intuitive, semantics-driven interaction in generative design tools.
\end{abstract}

\keywords{creativity support tool, visual exploration, generative AI}

\begin{teaserfigure}
  \centering
  \includegraphics[width=.97\textwidth]{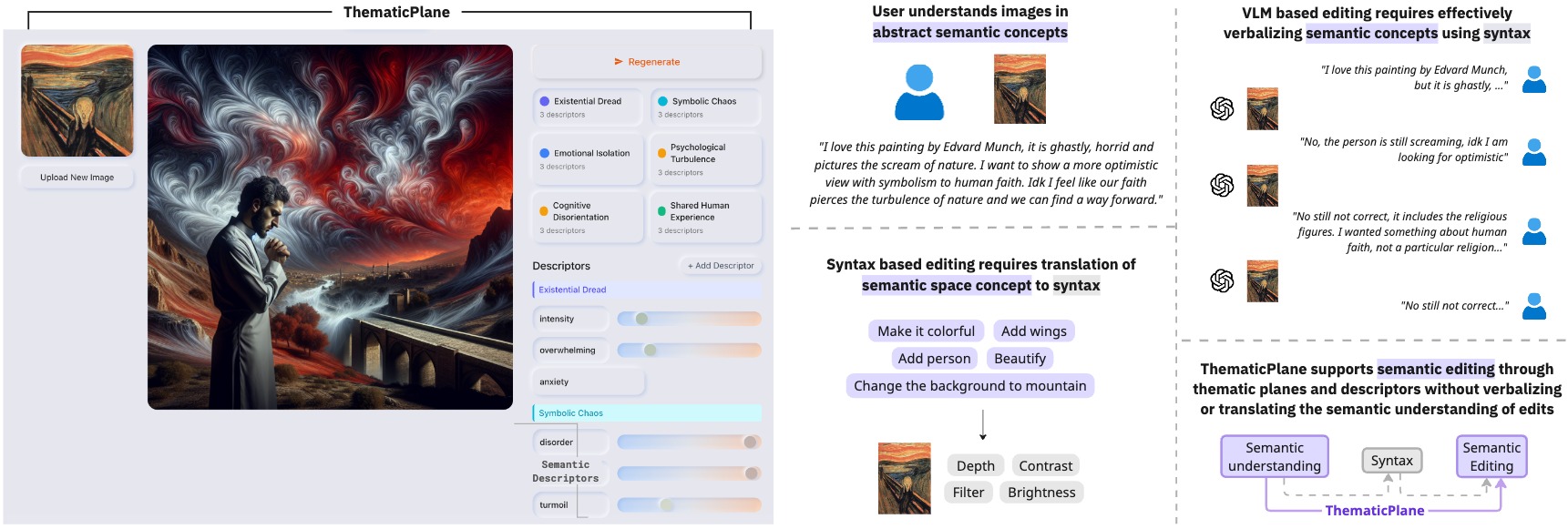}
  \caption{Overview of \system{}. Unlike conventional prompting, \system{} supports the navigation and manipulation of high-level semantic concepts within a thematic plan, allowing for tacit editing of the image aligned with the user's understanding of the semantic concept of the image.}
  \label{fig:teaser}
\end{teaserfigure}

\maketitle

\section{Introduction}

The proliferation of generative AI has fundamentally transformed creative expression in various modalities (\eg{}, text, images, video, code, etc.), enabling users to synthesize artifacts with unprecedented ease (\eg{}, via tools like ChatGPT or RunwayML). For instance, individuals without expertise in visual editing software can now generate or modify images through simple prompts, enabling lay users to construct sophisticated outputs~\cite{meng2023locatingeditingfactualassociations}. Nonetheless, achieving precise, intent-aligned results still demands technical knowledge (\eg{}, manually crafting detailed prompts, understanding the underlying model behaviors), hindering non-expert users~\cite{shi2024hcicentricsurveytaxonomyhumangenerativeai}.

\begin{figure}[b!]
    \centering
    \includegraphics[width=.8\linewidth]{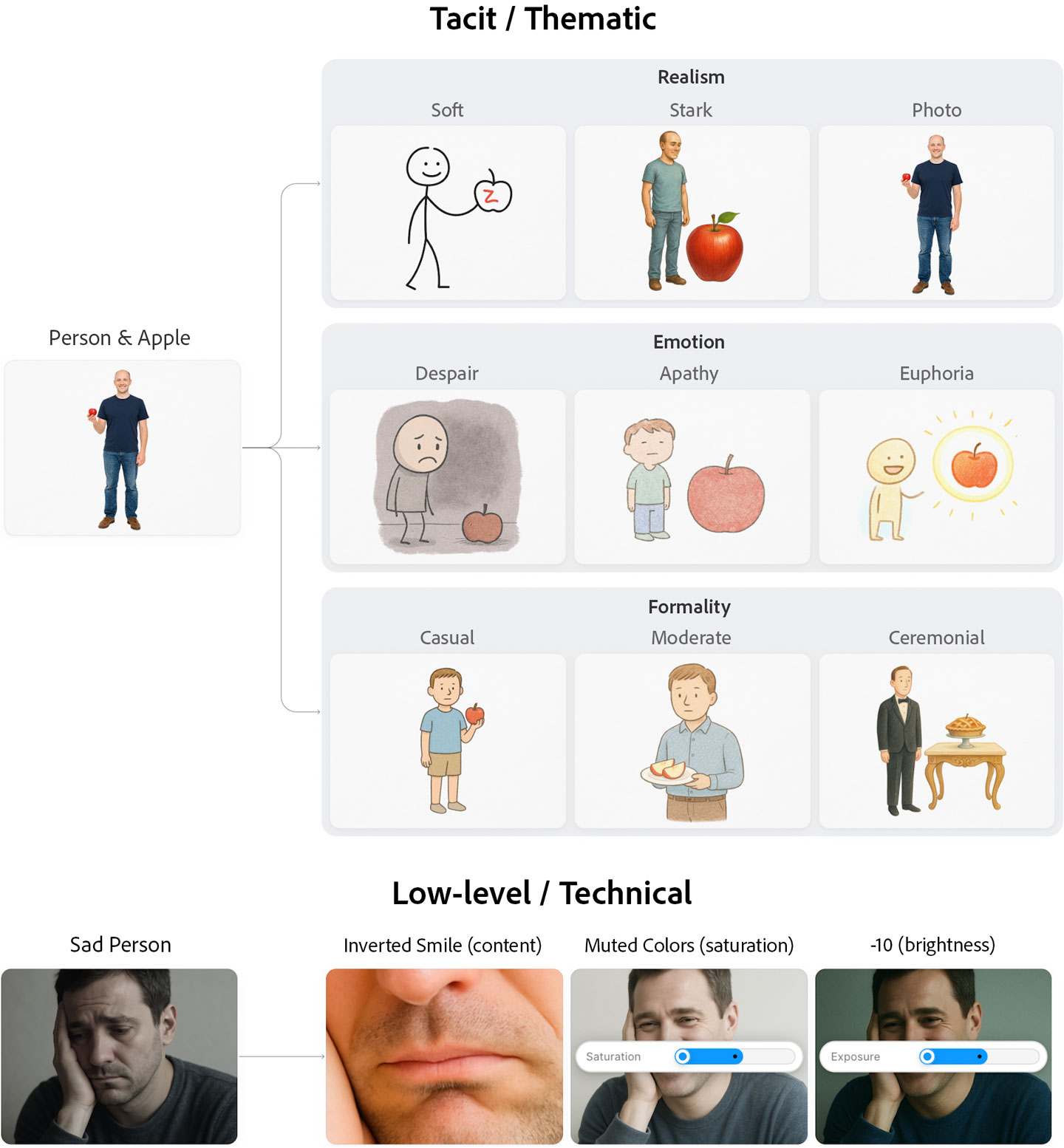}
    \caption{Visualization of tacit/thematic and low-level/technical edits in images.}
    \Description{}
    \label{fig:2}
\end{figure}

Previous works have proposed systems where users can change low-level image parameters (\eg{} with augmented prompt engineering scaffoldings~\cite{10.1145/3613904.3642803, gandikota2023conceptslidersloraadaptors}). However, users seldom conceptualize desired changes in terms of such low-level technical parameters, layers, or explicit instructions. Instead, they reason in high-level semantic terms, like aspiring to make an image ‘warmer,' ‘more dramatic,' or ‘friendlier'~\cite{article, 10.1145/3397481.3450670}. This semantic gap as visualized in \autoref{fig:2}--between users' tacit, thematic intent and systems' affordances--has spurred recent HCI research into intuitive editing paradigms, including keyword recombinations~\cite{choi2024creativeconnectsupportingreferencerecombination}, reference image-based modifications~\cite{son2024genquerysupportingexpressivevisual}, and interactive latent space explorations~\cite{oppenlaender2024promptingaiartinvestigation, 10.1145/3708359.3712150}.

Despite these, existing tools predominantly require users to externalize their vision through prescribed mediums, such as textual descriptions or reference images. This ‘externalization assumption' presumes users can readily translate abstract ideas into concrete formats, which may not align with natural cognitive processes of creative ideation and refinement~\cite{article, 10.1145/985692.985782, 10.1145/3706599.3720189}. As a result, workflows can become rigid, limiting iterative discovery in generative tasks. 

To address this, we introduce \system{}, a system that supports interactions within thematic design planes, which scaffolds the representation of multidimensional space for exploring and lets users manipulate thematic attributes of visual content. It enables users to ‘ground' and navigate high-level semantic concepts, such as mood, style, narrative, or emotional tone, in an exploratory fashion. This plane serves as a scaffolding layer, connecting users' tacit intent with system capabilities with a ‘plane of existence' for themes, drawing on semantic interaction paradigms to support creativity without externalization barriers.
\subsection{System Prototype}
\system{} implements the thematic design plane as an interactive framework adapted from the Circumplex Model~\cite{russell1980circumplex} for semantic image editing, enabling users to explore and manipulate high-level thematic attributes without explicit low-level inputs.
Given an input image by the user, GPT-4o extracts keywords, which are then refined by removing object-based descriptors and retaining only thematic elements like mood or style. Next, we generate 12 thematic perturbations for each theme, assigning them to left or right directions along a semantic axis. Embeddings are computed using DINOv2~\cite{oquab2024dinov2learningrobustvisual} for the input image and its aligned text encoder~\cite{jose2025dinov2} for textual themes, ranked by cosine similarity. Ranked descriptors are then injected into prompts for the image generation model (\ie{}, Imagen 3~\cite{imagenteamgoogle2024imagen3}) to produce transformed images. After each regeneration, users can choose to make the new image the primary reference to continue editing. 

\section{Exploratory Study}
We recruited 6 participants (4 women, 2 men), aged 19--31, through a corporate Slack channel. Participants had prior experience with image generation tools (\eg{}, Midjourney~\cite{midjourney_website}). They were first provided 10 minutes to explore the system, using their own images or study provided examples. We then designed three image transformation tasks requiring participants to modify source images to match target images. Each participant completed tasks across three conditions: (1) baseline ChatGPT constrained to image generation only, (2) ChatGPT with full features, and (3) our system---with 5 minutes per condition, followed by a semi-structured interview and survey. Each participant received \$20 for their participation.
\section{Discussion and Future Work}

\noindent \textit{\textbf{Aligning interfaces to the human creativity process.}} Our study revealed two emergent behaviors---open-ended discovery (\ie{}, divergent exploration) and goal-based editing (\ie{}, convergent refinement)~\cite{WILLEMSEN2023101375, guilford1959three}. Although participants using \system{} did not always anticipate the resulting images ($M = 3.43$, $SD = 1.27$), they reported high satisfaction with the outcomes ($M = 5.86$, $SD = 0.90$). Participants noted that non-linear generations either spurred intentional shifts in design direction during divergent tasks (P3) or enabled iteration in a space where themes amalgamated into cohesive affordances not characterized by a single plane (P6).\\[-2mm]

\noindent \textit{\textbf{Different perceptions of thematic planes.}} We observed that participants grounded themselves in themes based on their own understanding, enabling them to interpolate and navigate the rest of the thematic design plane; however, such understanding varied across participants, leading to differing expectations during exploration. Specifically, some participants (P1, P3, P7) anticipated linear mappings between themes and outputs, while others (P2, P4, P5) encountered unclear connections between the plane and initial themes, prompting a desire for explainable methods to comprehend steering mechanisms, rather than merely executing them~\cite{Bajaj_2024, 10.1145/3613904.3642133}.\\[-2mm]



\noindent \textit{\textbf{Interactive visualizations of themes.}} Despite the steerability it offers, our interaction method currently provides a single axis for users to steer generative outputs based on an initial thematic attribute, which may restrict the exploration of interconnected semantic dimensions \cite{10.1145/3613904.3642133}. Future work could advance this to multidimensional representations, such as interactive planes or graphs, enabling users to simultaneously manipulate multiple attributes (\eg{}, mood and style) through gestures like dragging or clustering for more nuanced control over latent space trajectories~\cite{10.1145/3613904.3642400}.\\[-2mm]

\noindent \textit{\textbf{Steerable model generations for images.}} Building on the visual representation of steerable dimensions, these should reflect the technical methods for navigating these spaces, extending text-based approaches like LM-Steer~\cite{han-etal-2024-word} to incorporate vector-based steering in multimodal generative systems~\cite{jahanian2020steerabilitygenerativeadversarialnetworks, esser2021tamingtransformershighresolutionimage} to create hybrid navigation mechanisms that dynamically adjust trajectories across semantic attributes during inference.

\begin{acks}
We are grateful to all participants who took part in our studies. We also wish to thank Anh Truong, Jaewook Lee, and Aditya Gunturu for their thoughtful feedback to the early development of this work.

\end{acks}

\bibliographystyle{ACM-Reference-Format}
\bibliography{paper-bibliography}


\begin{thebibliography}{25}


\ifx \showCODEN    \undefined \def \showCODEN     #1{\unskip}     \fi
\ifx \showISBNx    \undefined \def \showISBNx     #1{\unskip}     \fi
\ifx \showISBNxiii \undefined \def \showISBNxiii  #1{\unskip}     \fi
\ifx \showISSN     \undefined \def \showISSN      #1{\unskip}     \fi
\ifx \showLCCN     \undefined \def \showLCCN      #1{\unskip}     \fi
\ifx \shownote     \undefined \def \shownote      #1{#1}          \fi
\ifx \showarticletitle \undefined \def \showarticletitle #1{#1}   \fi
\ifx \showURL      \undefined \def \showURL       {\relax}        \fi
\providecommand\bibfield[2]{#2}
\providecommand\bibinfo[2]{#2}
\providecommand\natexlab[1]{#1}
\providecommand\showeprint[2][]{arXiv:#2}

\bibitem[Adamkiewicz et~al\mbox{.}(2025)]%
        {10.1145/3708359.3712150}
\bibfield{author}{\bibinfo{person}{Krzysztof Adamkiewicz}, \bibinfo{person}{Pawe{\l}~W Wo{\'z}niak}, \bibinfo{person}{Julia Dominiak}, \bibinfo{person}{Andrzej Romanowski}, \bibinfo{person}{Jakob Karolus}, {and} \bibinfo{person}{Stanislav Frolov}.} \bibinfo{year}{2025}\natexlab{}.
\newblock \showarticletitle{PromptMap: An Alternative Interaction Style for AI-Based Image Generation}. In \bibinfo{booktitle}{\emph{Proceedings of the 30th International Conference on Intelligent User Interfaces}}. \bibinfo{pages}{1162--1176}.
\newblock
\href{https://doi.org/10.1145/3708359.3712150}{doi:\nolinkurl{10.1145/3708359.3712150}}


\bibitem[Bajaj et~al\mbox{.}(2024)]%
        {Bajaj_2024}
\bibfield{author}{\bibinfo{person}{Goonmeet Bajaj}, \bibinfo{person}{Valerie~L Shalin}, \bibinfo{person}{Srinivasan Parthasarathy}, {and} \bibinfo{person}{Amit Sheth}.} \bibinfo{year}{2024}\natexlab{}.
\newblock \showarticletitle{Grounding from an AI and cognitive science lens}.
\newblock \bibinfo{journal}{\emph{IEEE Intelligent Systems}} \bibinfo{volume}{39}, \bibinfo{number}{2} (\bibinfo{year}{2024}), \bibinfo{pages}{66--71}.
\newblock
\href{https://doi.org/10.1109/mis.2024.3366669}{doi:\nolinkurl{10.1109/mis.2024.3366669}}


\bibitem[Baldridge et~al\mbox{.}(2024)]%
        {imagenteamgoogle2024imagen3}
\bibfield{author}{\bibinfo{person}{Jason Baldridge}, \bibinfo{person}{Jakob Bauer}, \bibinfo{person}{Mukul Bhutani}, \bibinfo{person}{Nicole Brichtova}, \bibinfo{person}{Andrew Bunner}, \bibinfo{person}{Lluis Castrejon}, \bibinfo{person}{Kelvin Chan}, \bibinfo{person}{Yichang Chen}, \bibinfo{person}{Sander Dieleman}, \bibinfo{person}{Yuqing Du}, {et~al\mbox{.}}} \bibinfo{year}{2024}\natexlab{}.
\newblock \showarticletitle{Imagen 3}.
\newblock \bibinfo{journal}{\emph{arXiv preprint arXiv:2408.07009}} (\bibinfo{year}{2024}).
\newblock
\urldef\tempurl%
\url{https://arxiv.org/abs/2408.07009}
\showURL{%
\tempurl}


\bibitem[Bian and North(2021)]%
        {10.1145/3397481.3450670}
\bibfield{author}{\bibinfo{person}{Yali Bian} {and} \bibinfo{person}{Chris North}.} \bibinfo{year}{2021}\natexlab{}.
\newblock \showarticletitle{Deepsi: Interactive deep learning for semantic interaction}. In \bibinfo{booktitle}{\emph{Proceedings of the 26th International Conference on Intelligent User Interfaces}}. \bibinfo{pages}{197--207}.
\newblock
\urldef\tempurl%
\url{https://doi.org/10.1145/3397481.3450670}
\showURL{%
\tempurl}


\bibitem[Choi et~al\mbox{.}(2024)]%
        {choi2024creativeconnectsupportingreferencerecombination}
\bibfield{author}{\bibinfo{person}{DaEun Choi}, \bibinfo{person}{Sumin Hong}, \bibinfo{person}{Jeongeon Park}, \bibinfo{person}{John Joon~Young Chung}, {and} \bibinfo{person}{Juho Kim}.} \bibinfo{year}{2024}\natexlab{}.
\newblock \showarticletitle{CreativeConnect: Supporting Reference Recombination for Graphic Design Ideation with Generative AI}. In \bibinfo{booktitle}{\emph{Proceedings of the 2024 CHI Conference on Human Factors in Computing Systems}}. \bibinfo{pages}{1--25}.
\newblock
\href{https://doi.org/10.1145/3613904.3642794}{doi:\nolinkurl{10.1145/3613904.3642794}}


\bibitem[Choi et~al\mbox{.}(2025)]%
        {10.1145/3706599.3720189}
\bibfield{author}{\bibinfo{person}{DaEun Choi}, \bibinfo{person}{Kihoon Son}, \bibinfo{person}{HyunJoon Jung}, {and} \bibinfo{person}{Juho Kim}.} \bibinfo{year}{2025}\natexlab{}.
\newblock \showarticletitle{Expandora: Broadening Design Exploration with Text-to-Image Model}. In \bibinfo{booktitle}{\emph{Proceedings of the Extended Abstracts of the CHI Conference on Human Factors in Computing Systems}}. \bibinfo{pages}{1--10}.
\newblock
\href{https://doi.org/10.1145/3706599.3720189}{doi:\nolinkurl{10.1145/3706599.3720189}}


\bibitem[Endert(2014)]%
        {article}
\bibfield{author}{\bibinfo{person}{Alex Endert}.} \bibinfo{year}{2014}\natexlab{}.
\newblock \showarticletitle{Semantic Interaction for Visual Analytics: Toward Coupling Cognition and Computation}.
\newblock \bibinfo{journal}{\emph{IEEE Computer Graphics and Applications}} \bibinfo{volume}{34}, \bibinfo{number}{4} (\bibinfo{year}{2014}), \bibinfo{pages}{8--15}.
\newblock
\href{https://doi.org/10.1109/MCG.2014.73}{doi:\nolinkurl{10.1109/MCG.2014.73}}


\bibitem[Esser et~al\mbox{.}(2021)]%
        {esser2021tamingtransformershighresolutionimage}
\bibfield{author}{\bibinfo{person}{Patrick Esser}, \bibinfo{person}{Robin Rombach}, {and} \bibinfo{person}{Bjorn Ommer}.} \bibinfo{year}{2021}\natexlab{}.
\newblock \showarticletitle{Taming Transformers for High-Resolution Image Synthesis}. In \bibinfo{booktitle}{\emph{Proceedings of the IEEE/CVF Conference on Computer Vision and Pattern Recognition}}. \bibinfo{pages}{12873--12883}.
\newblock
\href{https://doi.org/10.1109/CVPR46437.2021.01268}{doi:\nolinkurl{10.1109/CVPR46437.2021.01268}}


\bibitem[Gandikota et~al\mbox{.}(2024)]%
        {gandikota2023conceptslidersloraadaptors}
\bibfield{author}{\bibinfo{person}{Rohit Gandikota}, \bibinfo{person}{Joanna Materzy{\'n}ska}, \bibinfo{person}{Tingrui Zhou}, \bibinfo{person}{Antonio Torralba}, {and} \bibinfo{person}{David Bau}.} \bibinfo{year}{2024}\natexlab{}.
\newblock \showarticletitle{Concept Sliders: LoRA Adaptors for Precise Control in Diffusion Models}. In \bibinfo{booktitle}{\emph{European Conference on Computer Vision}}. Springer, \bibinfo{pages}{172--188}.
\newblock
\href{https://doi.org/10.1007/978-3-031-73661-2_10}{doi:\nolinkurl{10.1007/978-3-031-73661-2_10}}


\bibitem[Guilford(1959)]%
        {guilford1959three}
\bibfield{author}{\bibinfo{person}{J.P. Guilford}.} \bibinfo{year}{1959}\natexlab{}.
\newblock \showarticletitle{Three Faces of Intellect}.
\newblock \bibinfo{journal}{\emph{American Psychologist}} \bibinfo{volume}{14}, \bibinfo{number}{8} (\bibinfo{year}{1959}), \bibinfo{pages}{469--479}.
\newblock
\href{https://doi.org/10.1037/h0046827}{doi:\nolinkurl{10.1037/h0046827}}


\bibitem[Han et~al\mbox{.}(2024b)]%
        {han-etal-2024-word}
\bibfield{author}{\bibinfo{person}{Chi Han}, \bibinfo{person}{Jialiang Xu}, \bibinfo{person}{Manling Li}, \bibinfo{person}{Yi Fung}, \bibinfo{person}{Chenkai Sun}, \bibinfo{person}{Nan Jiang}, \bibinfo{person}{Tarek Abdelzaher}, {and} \bibinfo{person}{Heng Ji}.} \bibinfo{year}{2024}\natexlab{b}.
\newblock \showarticletitle{Word Embeddings Are Steers for Language Models}. In \bibinfo{booktitle}{\emph{Proceedings of the 62nd Annual Meeting of the Association for Computational Linguistics}}. \bibinfo{pages}{16410--16430}.
\newblock
\href{https://doi.org/10.18653/v1/2024.acl-long.864}{doi:\nolinkurl{10.18653/v1/2024.acl-long.864}}


\bibitem[Han et~al\mbox{.}(2024a)]%
        {10.1145/3613904.3642133}
\bibfield{author}{\bibinfo{person}{Yuanning Han}, \bibinfo{person}{Ziyi Qiu}, \bibinfo{person}{Jiale Cheng}, {and} \bibinfo{person}{Ray Lc}.} \bibinfo{year}{2024}\natexlab{a}.
\newblock \showarticletitle{When Teams Embrace AI: Human Collaboration Strategies in Generative Prompting in a Creative Design Task}. In \bibinfo{booktitle}{\emph{Proceedings of the 2024 CHI Conference on Human Factors in Computing Systems}}. \bibinfo{pages}{1--14}.
\newblock
\href{https://doi.org/10.1145/3613904.3642133}{doi:\nolinkurl{10.1145/3613904.3642133}}


\bibitem[Inc.(2025)]%
        {midjourney_website}
\bibfield{author}{\bibinfo{person}{Midjourney Inc.}} \bibinfo{year}{2025}\natexlab{}.
\newblock \bibinfo{title}{Midjourney}.
\newblock \bibinfo{howpublished}{\url{https://www.midjourney.com/}}.
\newblock


\bibitem[Jahanian et~al\mbox{.}(2019)]%
        {jahanian2020steerabilitygenerativeadversarialnetworks}
\bibfield{author}{\bibinfo{person}{Ali Jahanian}, \bibinfo{person}{Lucy Chai}, {and} \bibinfo{person}{Phillip Isola}.} \bibinfo{year}{2019}\natexlab{}.
\newblock \showarticletitle{On the "steerability" of generative adversarial networks}.
\newblock \bibinfo{journal}{\emph{arXiv preprint arXiv:1907.07171}} (\bibinfo{year}{2019}).
\newblock
\urldef\tempurl%
\url{https://arxiv.org/abs/1907.07171}
\showURL{%
\tempurl}


\bibitem[Jose et~al\mbox{.}(2025)]%
        {jose2025dinov2}
\bibfield{author}{\bibinfo{person}{Cijo Jose}, \bibinfo{person}{Th{\'e}o Moutakanni}, \bibinfo{person}{Dahyun Kang}, \bibinfo{person}{Federico Baldassarre}, \bibinfo{person}{Timoth{\'e}e Darcet}, \bibinfo{person}{Hu Xu}, \bibinfo{person}{Daniel Li}, \bibinfo{person}{Marc Szafraniec}, \bibinfo{person}{Micha{\"e}l Ramamonjisoa}, \bibinfo{person}{Maxime Oquab}, {et~al\mbox{.}}} \bibinfo{year}{2025}\natexlab{}.
\newblock \showarticletitle{DINOv2 Meets Text: A Unified Framework for Image- and Pixel-Level Vision-Language Alignment}. In \bibinfo{booktitle}{\emph{Proceedings of the Computer Vision and Pattern Recognition Conference}}. \bibinfo{pages}{24905--24916}.
\newblock


\bibitem[Meng et~al\mbox{.}(2022)]%
        {meng2023locatingeditingfactualassociations}
\bibfield{author}{\bibinfo{person}{Kevin Meng}, \bibinfo{person}{David Bau}, \bibinfo{person}{Alex Andonian}, {and} \bibinfo{person}{Yonatan Belinkov}.} \bibinfo{year}{2022}\natexlab{}.
\newblock \showarticletitle{Locating and Editing Factual Associations in GPT}.
\newblock \bibinfo{journal}{\emph{Advances in Neural Information Processing Systems}}  \bibinfo{volume}{35} (\bibinfo{year}{2022}), \bibinfo{pages}{17359--17372}.
\newblock


\bibitem[Oppenlaender et~al\mbox{.}(2024)]%
        {oppenlaender2024promptingaiartinvestigation}
\bibfield{author}{\bibinfo{person}{Jonas Oppenlaender}, \bibinfo{person}{Rhema Linder}, {and} \bibinfo{person}{Johanna Silvennoinen}.} \bibinfo{year}{2024}\natexlab{}.
\newblock \showarticletitle{Prompting AI Art: An Investigation into the Creative Skill of Prompt Engineering}.
\newblock \bibinfo{journal}{\emph{International Journal of Human–Computer Interaction}} (\bibinfo{year}{2024}), \bibinfo{pages}{1--23}.
\newblock
\href{https://doi.org/10.1080/10447318.2024.2431761}{doi:\nolinkurl{10.1080/10447318.2024.2431761}}


\bibitem[Oquab et~al\mbox{.}(2023)]%
        {oquab2024dinov2learningrobustvisual}
\bibfield{author}{\bibinfo{person}{Maxime Oquab}, \bibinfo{person}{Timoth{\'e}e Darcet}, \bibinfo{person}{Th{\'e}o Moutakanni}, \bibinfo{person}{Huy Vo}, \bibinfo{person}{Marc Szafraniec}, \bibinfo{person}{Vasil Khalidov}, \bibinfo{person}{Pierre Fernandez}, \bibinfo{person}{Daniel Haziza}, \bibinfo{person}{Francisco Massa}, \bibinfo{person}{Alaaeldin El-Nouby}, {et~al\mbox{.}}} \bibinfo{year}{2023}\natexlab{}.
\newblock \showarticletitle{DINOv2: Learning Robust Visual Features without Supervision}.
\newblock \bibinfo{journal}{\emph{arXiv preprint arXiv:2304.07193}} (\bibinfo{year}{2023}).
\newblock
\urldef\tempurl%
\url{https://arxiv.org/abs/2304.07193}
\showURL{%
\tempurl}


\bibitem[Russell(1980)]%
        {russell1980circumplex}
\bibfield{author}{\bibinfo{person}{James~A Russell}.} \bibinfo{year}{1980}\natexlab{}.
\newblock \showarticletitle{A circumplex model of affect.}
\newblock \bibinfo{journal}{\emph{Journal of personality and social psychology}} \bibinfo{volume}{39}, \bibinfo{number}{6} (\bibinfo{year}{1980}), \bibinfo{pages}{1161}.
\newblock


\bibitem[Shi et~al\mbox{.}(2023)]%
        {shi2024hcicentricsurveytaxonomyhumangenerativeai}
\bibfield{author}{\bibinfo{person}{Jingyu Shi}, \bibinfo{person}{Rahul Jain}, \bibinfo{person}{Hyungjun Doh}, \bibinfo{person}{Ryo Suzuki}, {and} \bibinfo{person}{Karthik Ramani}.} \bibinfo{year}{2023}\natexlab{}.
\newblock \showarticletitle{An HCI-Centric Survey and Taxonomy of Human-Generative-AI Interactions}.
\newblock \bibinfo{journal}{\emph{arXiv preprint arXiv:2310.07127}} (\bibinfo{year}{2023}).
\newblock
\urldef\tempurl%
\url{https://arxiv.org/abs/2310.07127}
\showURL{%
\tempurl}


\bibitem[Son et~al\mbox{.}(2024)]%
        {son2024genquerysupportingexpressivevisual}
\bibfield{author}{\bibinfo{person}{Kihoon Son}, \bibinfo{person}{DaEun Choi}, \bibinfo{person}{Tae~Soo Kim}, \bibinfo{person}{Young-Ho Kim}, {and} \bibinfo{person}{Juho Kim}.} \bibinfo{year}{2024}\natexlab{}.
\newblock \showarticletitle{GenQuery: Supporting Expressive Visual Search with Generative Models}. In \bibinfo{booktitle}{\emph{Proceedings of the 2024 CHI Conference on Human Factors in Computing Systems}}. \bibinfo{pages}{1--19}.
\newblock
\href{https://doi.org/10.1145/3613904.3642847}{doi:\nolinkurl{10.1145/3613904.3642847}}


\bibitem[Suh et~al\mbox{.}(2024)]%
        {10.1145/3613904.3642400}
\bibfield{author}{\bibinfo{person}{Sangho Suh}, \bibinfo{person}{Meng Chen}, \bibinfo{person}{Bryan Min}, \bibinfo{person}{Toby Jia-Jun Li}, {and} \bibinfo{person}{Haijun Xia}.} \bibinfo{year}{2024}\natexlab{}.
\newblock \showarticletitle{Luminate: Structured Generation and Exploration of Design Space with Large Language Models for Human-AI Co-Creation}. In \bibinfo{booktitle}{\emph{Proceedings of the 2024 CHI Conference on Human Factors in Computing Systems}}. \bibinfo{pages}{1--26}.
\newblock
\href{https://doi.org/10.1145/3613904.3642400}{doi:\nolinkurl{10.1145/3613904.3642400}}


\bibitem[Terry et~al\mbox{.}(2004)]%
        {10.1145/985692.985782}
\bibfield{author}{\bibinfo{person}{Michael Terry}, \bibinfo{person}{Elizabeth~D Mynatt}, \bibinfo{person}{Kumiyo Nakakoji}, {and} \bibinfo{person}{Yasuhiro Yamamoto}.} \bibinfo{year}{2004}\natexlab{}.
\newblock \showarticletitle{Variation in Element and Action: Supporting Simultaneous Development of Alternative Solutions}. In \bibinfo{booktitle}{\emph{Proceedings of the SIGCHI Conference on Human Factors in Computing Systems}}. \bibinfo{pages}{711--718}.
\newblock
\href{https://doi.org/10.1145/985692.985782}{doi:\nolinkurl{10.1145/985692.985782}}


\bibitem[Wang et~al\mbox{.}(2024)]%
        {10.1145/3613904.3642803}
\bibfield{author}{\bibinfo{person}{Zhijie Wang}, \bibinfo{person}{Yuheng Huang}, \bibinfo{person}{Da Song}, \bibinfo{person}{Lei Ma}, {and} \bibinfo{person}{Tianyi Zhang}.} \bibinfo{year}{2024}\natexlab{}.
\newblock \showarticletitle{PromptCharm: Text-to-Image Generation through Multi-modal Prompting and Refinement}. In \bibinfo{booktitle}{\emph{Proceedings of the 2024 CHI Conference on Human Factors in Computing Systems}}. \bibinfo{pages}{1--21}.
\newblock
\href{https://doi.org/10.1145/3613904.3642803}{doi:\nolinkurl{10.1145/3613904.3642803}}


\bibitem[Willemsen et~al\mbox{.}(2023)]%
        {WILLEMSEN2023101375}
\bibfield{author}{\bibinfo{person}{Robin~H Willemsen}, \bibinfo{person}{Isabelle~C de Vink}, \bibinfo{person}{Evelyn~H Kroesbergen}, {and} \bibinfo{person}{Ard~W Lazonder}.} \bibinfo{year}{2023}\natexlab{}.
\newblock \showarticletitle{The role of creative thinking in children's scientific reasoning}.
\newblock \bibinfo{journal}{\emph{Thinking Skills and Creativity}}  \bibinfo{volume}{49} (\bibinfo{year}{2023}), \bibinfo{pages}{101375}.
\newblock
\href{https://doi.org/10.1016/j.tsc.2023.101375}{doi:\nolinkurl{10.1016/j.tsc.2023.101375}}


\end{thebibliography}

\appendix

\end{document}